\documentclass[12pt,a4paper]{article}

\usepackage[english]{babel}

\usepackage{epsfig}
\usepackage{dcolumn}
\usepackage{amsmath}
\usepackage{cite}
\usepackage{changebar}

\makeatletter \@addtoreset{equation}{section}
\makeatother 
\renewcommand{\thefootnote}{\alph{footnote}}

\newcommand{\Tr} {\mbox{Tr}}

\begin{document}
\thispagestyle{empty}
\hbox{}
 \mbox{} \hfill \hspace{1.0cm}
         \today 

 \mbox{} \hfill BI-TP 2003/14\hfill

\begin{center}
\vspace*{1.0cm}
\renewcommand{\thefootnote}{\fnsymbol{footnote}}
{\huge \bf  ENTROPY FOR $SU(3)_c$}\\  
{\huge \bf QUARK STATES}\\

\vspace*{1.0cm}
{\large David E. Miller$^{1,2}$}
\\
\vspace*{1.0cm}
${}^1$ Fakult\"at f\"ur Physik, Universit\"at Bielefeld, Postfach 100131,\\ 
D-33501 Bielefeld, Germany\footnote{email: dmiller@physik.uni-bielefeld.de} \\
${}^2$ Department of Physics, Pennsylvania State University,
Hazleton Campus,\\
Hazleton, Pennsylvania 18201, USA 
\footnote{permanent address, email: om0@psu.edu} \\
\vspace*{2cm}
{\large \bf Abstract}
\end{center}

We discuss the quantum  state structure using the standard model
for three colored quarks in the fundamental representations of $SU(3)_c$
making up the singlet ground state of the hadrons. This allows us
to calculate a finite von Neumann entropy from the quantum reduced
density matrix, which we explicitly evaluate for the quarks in a
model for the meson and baryon states. Finally we look into the
general effects and implications of entanglement in the $SU(3)_c$
color space.

\vfill
\noindent PACS numbers:12.38Aw, 03.75.Ss, 12.39.-x, 05.30.-d
\newpage


\section{\bf Introduction}

\noindent
The well known theorem of Nernst, which is often times referred to as 
the Third Law of Thermodynamics, has the generally accepted interpretation
in the theory of gases that the entropy vanishes in the zero temperature
limit. Schr\"odinger~\cite{Schr} had pointed out long ago that when two 
states contribute to the ground state of a many particle
system that a finite constant term could appear in the entropy. In 
particular, for a system with $~2^N~$ states making up the ground state
of a system of N particles one should expect a ground state entropy of
$~N\ln{2}$. We can now understand his result in terms of the $~SU(2)~$
symmetry for the N particles. In this sense we should expect an internal
symmetry provided by the quantum structure to yield an entropy following 
the prescription of von Neumann~\cite{vNeuPer}.
~\\
~\\
\indent
     The  standard model has the color charge carried by the quarks as
the fundamental property of the strong nuclear interaction~\cite{DoGoHo}. 
In clear contrast to the other known charges the color charge cannot be 
easily isolated and separately measured. In nature it always appears as
part of selective states of the $SU(3)_c$, wherein the quarks and antiquarks
are placed in the fundamental $~\bf{3}~$ and antifundamental $~\bf{3^{*}}~$ 
representations of this group.  
These two representations together with the adjoint representation
of $SU(3)_c$ make up the symmetry structure of Quantum Chromodynamics(QCD)
~\cite{Mut}. The two main categories of strong interacting particles (hadrons)
are the mesons, which may be written as a product of the fundamental and the 
antifundamental representations $~\bf{3}~{\otimes}~\bf{3^{*}}~$ and the baryons,
which are a product of three fundamental representations 
$~\bf{3}~{\otimes}~\bf{3}~{\otimes}~\bf{3}~$.  Although the different quarks have
other properties like spin, electrical charge, mass as well as a very
special property called {\it{flavor}}, we shall not presently go into 
these aspects here~\cite{DoGoHo}. 
~\\
~\\
\indent 
     In this work we shall write the quark and antiquark color states as 
follows: $~|0\rangle~$,$~|1\rangle~$,$~|2\rangle~$ 
and$~|0^{*}\rangle~$,$~|1^{*}\rangle~$,$~|2^{*}\rangle~$. 
We shall use this notation to describe the orthonormal bases of the 
fundamental and the antifundamental representations of $SU(3)_c$ instead
of the more common color names. From these color states we can construct
a representation for the color hadronic wavefunctions-- in particular for
the singlet meson ${\Psi}_{M,s}$ and baryon ${\Psi}_{B,s}$ groundstates.
We also mention the construction for the eight density matrices for the 
color octet states of the mesons and baryons.
From these color wavefunctions we are able to construct the corresponding
density matrices ${\bf{\rho}}_{M,s}$ and ${\bf{\rho}}_{B,s}$ for the color
singlet states~\cite{LaLi}. In the following work we shall arrive at the
single quark reduced density matrix ${\bf{\rho}}_{q}$, which is of
particular interest in all further calculations. From ${\bf{\rho}}_{q}$
we can directly calculate the quantum entropy in the sense of von Neumann
~\cite{LaLi,vNeuPer}. The results of this calculation show a significant
contribution of order one to the entropy of the quarks in the hadronic
singlet and octet states. This value is given as a pure number without 
physical dimensions when we use the usual high energy units with 
$\hbar$, $c$ and Boltzmann's constant $k$ all set to the value one.
~\\
~\\
\indent 
     The further implications of these results can be brought together 
with some of our earlier work. After a discussion of the entropy density
we look at the thermodynamics of the quark states at very low temperatures,
which means at temperatures much lower than the lightest quark masses.
We rewrite the energy density equation in a form involving the trace of the 
energy momentum tensor. This allows us to include the vacuum condensates in the
low temperature limit. Next we relate this work to some earlier work involving a
finite number of baryons at finite temperatures. We discuss some recent work
involving the hadron resonance mass spectrum and its relationship to lattice QCD 
thermodynamics. At high temperatures the entropy density s(T) from gluons is
well known from SU(3) lattice simulations. Finally we mention some recent work
relating to the quantum mechanics of entangled states as it is often applied
to quantum information theory.


\section{\bf Singlet Quark Structure}

\indent
     The starting point for the ground state is the evaluation of the
density matrix~\cite{LaLi}
 for the singlet quark structure. Here we only consider
the color part of the wavefunctions ${\Psi}_{M,s}$ and ${\Psi}^{*}_{M,s}$ 
coming from the representation  $~\bf{3}~{\otimes}~\bf{3^{*}}~$ 
for the color singlet wavefunctions of the mesons,
\begin{equation}
 {\Psi}_{M,s}~=~{\frac{1}{\sqrt{3}}}(|0~0^*\rangle~+~
               |1~1^*\rangle~+~|2~2^*\rangle).
  \label{eq:mesonsing}
\end{equation}
and keeping the left to right order of the quark and antiquark 
for its conjugate wavefunction
\begin{equation}
 {\Psi}^{*}_{M,s}~=~{\frac{1}{\sqrt{3}}}(\langle0~0^* |~+~
                   \langle 1~1^*|~+~\langle 2~2^*|).
  \label{eq:mesonsingstar}
\end{equation}

Similarly we may write a wavefunction for the baryons ${\Psi}_{B,s}$ 
coming from the\\
 representation $~\bf{3}~{\otimes}~\bf{3}~{\otimes}~\bf{3}~$  
for the color singlet state of the baryons,
\begin{equation}
 {\Psi}_{B,s}~=~{\frac{1}{\sqrt{6}}}(|0~1~2\rangle~+~
               |1~2~0\rangle~+~|2~0~1\rangle~
               -~|0~2~1\rangle~-~|1~0~2\rangle
                ~-~|2~1~0\rangle ).
  \label{eq:baryonsing}
\end{equation}
The conjugate state wavefunction in the order of the tensor product 
for the baryons is given by
\begin{equation}
{\Psi}^{*}_{B,s}~=~~{\frac{1}{\sqrt{6}}}(\langle 0~1~2|~+~
               \langle 1~2~0|~+~\langle 2~0~1|~
              -~\langle 0~2~1|~-~\langle 1~0~2|
                ~-~\langle 2~1~0|).
  \label{eq:baryonsingstar}
\end{equation}
\noindent
We can now write down the density matrices $\bf{\rho}$ for the hadrons using
the direct product of ${\Psi}$ and ${\Psi}^{*}$. This gives for the color
singlet mesons and baryons the density matrices in the following forms:
\begin{equation}
{\bf{\rho}}_{M,s}~=~{\Psi}_{M,s} {\Psi}^{*}_{M,s}
  \label{eq:denmeson}
\end{equation}
and
\begin{equation}
{\bf{\rho}}_{B,s}~=~{\Psi}_{B,s} {\Psi}^{*}_{B,s} .  
  \label{eq:denbaryon}
\end{equation}
~\\
\indent
     Up until now we have only considered the hadronic states as being
made out of the quark states. The resulting density matrices are for the 
hadrons $\it{pure}$ states~\cite{LaLi}. However, for the
quarks we look at the one quark reduced density matrices, which give the
statistical state of the individual quark within the hadron. In order to
get the reduced density matrices for the mesons, we project out all the
antiquark states $~\langle i^*|~$and $~|j^*\rangle~$ by using the
orthonormality and the completeness properties. Similarly for the 
baryons we project onto the other two quark states resulting in two
contributions for each color. Thus both the meson and the baryon
reduced density matrices for the quark states take on the same form:
\begin{equation}
\label{eq:reddenmes}
{\bf{\rho}}_{q}~=~{\frac{1}{3}}(|0\rangle \langle 0|~+~
                      |1 \rangle \langle1|~+~|2\rangle \langle 2|).
\end{equation}
\noindent
This is the reduced density matrix for the quarks in the color singlet state.
It yields a completely mixed state where each color contribution has
the same eigenvalue $\lambda_i$ equal to the value $~1/3~$. The reduced density
matrices can also be calculated for each quark state in the octet
representations. A more detailed discussion will appear in a later work.



\section{\bf Entropy for Quark States} 

We can calculate the entropy $~S~$ of the quantum states using the
prescription of von Neumann~\cite{LaLi,vNeuPer}, which makes direct
use of the density matrix $~\bf{\rho}$. It is simply written as
\begin{equation}
\label{eq:entropy}
                   S~=~-{\Tr(~\bf{\rho}}~{\ln{\bf{\rho}})},
\end{equation}
where the trace "Tr" is taken over the quantum states. When, as is
presently the case, the eigenvectors are known for $~\rho~$, we may
write this form of the entropy in terms of the eigenvalues $~{\lambda}_i$
as follows: 
\begin{equation}
\label{eq:enteig}
                   S~=~-{{\lambda}_i{\ln{\lambda}_i}}.
\end{equation}
It is obviously important to have positive eigenvalues. For a zero
eigenvalue we use the fact that $x{\ln x}$ vanishes in the small x limit.
Then for the density matrix $~\rho~$ we may interpret ${\lambda}_i$ as the
probablitiy of the state $~i~$ or $~p_i~$. This meaning demands that
$~0~<~p_i~\leq~1~$. Thus the orthonormality condition for the given states
results in the trace condition
\begin{equation}
\label{eq:tracecond}
                  \Tr~{\rho}~=~{\sum}_i p_i~=~1.
\end{equation}
This is a very important condition for the entropy.
~\\
\indent
     We now apply these definitions to the entropy for the quark states.
It is clear that the original hadron states are pure colorless states
which posess zero entropy. For the meson it is immediately obvious since
each colored quark state has the opposing colored antiquark state for
the resulting colorless singlet state. The sum of all the cycles determine
the colorlessness of the baryonsinglet state thereby giving no entropy.
However, the reduced density matrix for the individual quarks (antiquarks)
$~{\rho}_q~$ or $~{\rho}_{\bar{q}}~$ has a finite entropy. For $SU(3)_c$
all the eigenvalues $~{\lambda}_i~$ in Equation (2.4) have the same value 1/3.
Thus we find for all the quarks (antiquarks) in  singlet states
\begin{equation}
\label{eq:entquark}
                  S_q~=~\ln 3.
\end{equation}
\indent
     As a further point we may draw a qualitative comparison of this result 
for the singlet state with the entropies of the quark octet states. 
The octet density matrices $~{\rho}_{o,i}~$ may
be constructed from the eight Gell-Mann matrices $~(\lambda)_i~$ with 
$~i~=~1,2,\cdots,8~$. The density matrix for each state is constructed by
using the properties of $~{\Psi}(\lambda)_i{\Psi}^{*}~$. From the reduced
density matrix, the first seven of these all give the same value for the 
entropy $~S_{o,i}$, since all of these states are constructed only from
the Pauli matrices. However, the eighth diagonal Gell-Mann matrix yields a
larger entropy $~S_{o,8}~$ from the fact that it\\ 
involves all three colors although not with equal weights as it was the case 
for the color singlet state. 
~\\
~\\
\indent
     Hereupon we may discuss the entropy in some more detail for the main examples 
of the colorless hadronic ground states-- the mesons and the baryons. 
As we have discussed above for the density matrix, 
all the mesons consist of a quark-antiquark pair bound together as a sum of
all the three colors. Since each single quark or antiquark state is equally
weighted in the reduced density matrix, therefore each state posesses equal 
probability of 1/3. Thus we easily get the entropy of $~\ln~3~$. The baryon has 
the doubly reduced density matrix for each single quark state appearing twice so 
that with the normalization factor of 1/6 the probability of each colored quark 
state is again 1/3, which yields the same result for the entropy, $~\ln~3~$. 
This value gives the maximal entropy for a completely  mixed state. We know that 
the singlet and the octet states make up a major contribution to all the hadronic
states.  
~\\
~\\
\indent
     It is the colored quark entropy density which physically distinguishes 
the thermodynamics of the baryons from the mesons. If we use the generally
accepted values for the root mean squared charge radius of the hadrons
~\cite{DoGoHo},  we take for the mesons (pions) $~\sqrt{\langle r_{M}^2 \rangle}~$
as $~0.66~\pm~0.02~fm$, while for the baryons (protons)~\cite{PDG} we use
$~\sqrt{\langle r_{B}^2 \rangle}~$ the value $~0.870~\pm~0.008~fm$. These
values of the charge radii are small when put on the nuclear size scale, where
we would generally expect sizes well over 1 fm. These mean charge radii give 
spherical charge volumes for the mesons ranging from $~1.098~fm^3~$ to
$~1.317~fm^3~$ or about $~1.20~fm^3~$ as the average mean volume.
Similarly for the baryons we find an average mean volume of $~2.76~fm^3~$.
The mean entropy density of the quarks in the singlet ground state of the 
mesons (pions) is given by
\begin{equation}
        s_{M}~=~{\frac{\ln~3}{1.20~fm^3}}~=~0.912~{\frac{1}{fm^3}} .
\label{eq:entdenmes}
\end{equation}
Similarly for the baryons in the singlet ground state we arrive at a mean 
value for the entropy density
\begin{equation}
        s_{B}~=~{\frac{\ln~3}{2.76~fm^3}}~=~0.398~{\frac{1}{fm^3}} .
\label{eq:entdenbar}
\end{equation}
These entropy densities represent the most probable distributions of quarks
in the given volume for the charged portions of the hadrons. In the next section
we shall discuss the relationship of these results to the thermodynamics of 
massive quark systems.
~\\
~\\
\indent
     As a last remark in this section on the meaning of this type of entropy
for the quantum ground state we should note that the effects of this type
appear in other systems with internal symmetries. In quantum spin chains
~\cite{LatRiVi}~ the effects of the ground state entanglement show
strong correlations in a block of L spins giving entropies proportional
to the logarithm of the size L in the various special cases of the
quantum Heisenberg model. These results are then related to the entropy
in a $~1~+~1~$dimensional conformal field theory.
Furthermore, one could, perhaps, 
extend these results  to a three state model like the $Z(3)$ symmetric
spin models or the extended Potts models~\cite{GaKa} to find 
analogous properties for the entropy in the low temperature limit.
\section{Thermodynamics for confined Quarks}
\noindent
We start with the form of the First Law of Thermodynamics for a quark
gluon system at very low temperatures T.
In terms of the densities we write
\begin{equation}
       s(T)T~=~\epsilon(T)~+~p(T)~-~\mu_q{n_q(T)},
\label{eq:thermden}
\end{equation}
\noindent
where $\mu_q$ is the quark chemical potential and $n_q(T)$ is
quark density distribution function.
We may rewrite this equation using the fact that the thermal
average of the trace of the energy-momentum tensor is written as
the equation of state~\cite{LaLi}
\begin{equation}
    \langle \Theta^{\mu}_{\mu}\rangle_T ~=~\epsilon(T)~-~3p(T),
\label{eq:enmomten}
\end{equation}
\noindent
where the sum is taken on the Lorentz indices $\mu$.
We now rewrite equation (4.1) as
\begin{equation}
    \langle \Theta^{\mu}_{\mu}\rangle_T ~=~s(T)T~-~4p(T)~+~\mu_q{n_q(T)} .
\label{eq:tempenmomten}
\end{equation}
In the limit that the temperature goes to zero we use $~p~=~-\mathcal{B}~$ 
where $\mathcal{B}$ is the bag constent~\cite{DoGoHo}. 
Clearly $~s(T)T~$ vanishes. 
Thus our equation of state in the low temperature limit the Fermi distribution
function $f(\mu_q,T)$ simply becomes a step function so that
\begin{equation}
    \langle \Theta^{\mu}_{\mu}\rangle_0 ~=~4\mathcal{B}~+~\mu_q{n_0}f(\mu_q) .
\label{eq:eqstatea}
\end{equation}
The trace of the energy-momentum tensor can be related to the gluon and quark
vacuum eapectation values arising from the operator product expansion
~\cite{DoGoHo, SVZ, Leut} taking the following form:
\begin{equation}
~\langle \Theta^{\mu}_{\mu}\rangle_0~=~ \langle G^2 \rangle_0
                                 ~+~m_q\langle\bar{\psi}_q{\psi}_q \rangle_0~
\label{eq:vaccond}
\end{equation}
Thus we have the important vacuum contributions of operator dimension four to
the equation of state, which are independent of the temperature.
~\\
~\\
\indent
     We choose as a simple special case that of the meson with the same light 
quarks surrounded by nuclear matter as an example for this investigation. 
Then we have from the Fermi statistics using the quark-antiquark symmetry 
$~\mu_{\bar q}~=~-\mu_q~$ and for the antiquark quantum density distribution
function $~n_{\bar{q}}(T)~=~n_0(1~-~f(T))~$, where $~n_0~$ is the quark 
density at zero temperature. Thus in the  limit of zero temperature we have simply
\begin{equation}
    \langle \Theta^{\mu}_{\mu}\rangle_0 ~=~4\mathcal{B}.
\label{eq:eqstateb}
\end{equation}
We know from Leutwyler's estimate~\cite{Leut} that the gluon condensate is about
 $~2GeV/fm^3~$ as well as the quark condensate for light quarks is almost
negligible. Thus we find that for this case the bag constant $~\mathcal{B}~$
is rather big . However, if we take T finite and small, the gluon condensate 
does not change much~\cite{Leut, BoMi}, but the entropy does the play a very 
important role. The equation of state becomes with $~2s_M~$ for both quarks
and antiquarks
\begin{equation}
    \langle \Theta^{\mu}_{\mu}\rangle_T ~=~(2s_M~+~s(T))T~-~4(p(T)~-~\mathcal{B})
                                          ~+~\mu_q n_0(2f(T)~-~1).
\label{eq:meseqnst}
\end{equation}
At very low temperatures we would expect that $~s(T)~$, $~p(T)~$ and
$n_q(T)$ all to remain insignificant so that the main change in the 
equation of state is the value of $~s_M~$. This effect could relate 
to a lowered bag constant or a raised chemical potential.
However, we presently do not know how much the finite
temperature changes the actual value of $~s_M~$.

\section{Discussions and Conclusions}
\noindent
We have calculated the entropy for a single quark in the color singlet ground state
of the hadrons. We saw that the singlet state is a completely mixed state with
the maximum value of the entropy given by $\ln 3$. In the theory of information
it is known that the completely mixed state is that of minimal information,
which is consistent with the idea of confinement. If we were to consider a gas
of N hadrons in the sense that Schr\"odinger~\cite{Schr} considered a gas of
two level atoms, we would generally expect an entropy of the form $N{\ln 3}$ 
for the color singlet ground state.
~\\
~\\
\indent
     In an earlier collaborative work~\cite{MiRe} we considered the
thermodynamical properties of a class of models with non-abelian internal
symmetries at finite temperature and baryon number. In this model we calculated
the entropy density for N quarks $s_N$ for fairly high temperatures and small
quark numbers N. We noted that the quantum effects became increasingly important
when we took larger quark number with lower temperatures and smaller volumes.
However, we did not then further investigate this issue.
~\\
~\\
\indent
     Further work on the transition from the hadronic phase to the quark-gluon
plasma has been recently carried out~\cite{KaReTa}.  Although the temperatures
are generally in the range  above 100MeV for comparison with the lattice data,
the hadron resonance gas model can be used for much lower temperatures. It is 
also useful to take into account the values~\cite{Boyd} for $~s(T)~$ from the 
pure $~SU(3)_c~$ lattice simulations~\cite{Eng3} as a comparison even though the
critical temperature $~T_c~$ is much higher at about 264MeV. However, the MILC 
two light quark flavors data~\cite{MILC97} shows quite different behavior 
at much lower temperatures beginning around 125MeV as does the four flavor 
data~\cite{BI97}. Furthermore, some very interesting three quark simulations 
have recently appeared~\cite{KLP}. 
~\\
~\\
\indent
     Finally we mention quantum information theory as another field of science
where similar techniques involving the density matrices and the corresponding
entropies are used. Although the difficulties arising from the quantum
mechanical structure present in the statistical physics have been recognized 
for many years~\cite{Schr,LaLi}, it is only quite recently that it has received
much attention outside these fields-- particularly in quantum information
theory~\cite{NieChu}. Here elaborate systems of codes are treated like quantum
spin states. Also further studies in mathematical physics relating to the 
entanglement problem itself have been recently carried out~\cite{BlaJaOl}.
Nevertheless, our objective here has been naturally provided by the entanglement
of the quark color structure as the extension of this approach to the three
basis states of the fundamental representations of $SU(3)_c$.
~\\
~\\
\medskip 
{\bf\Large Acknowledgements}
~\\
~\\
\medskip
The author would like to thank Philippe Blanchard, Frithjof Karsch, Krzysztof
Redlich,
Dieter Schildknecht and especially Abdelnasser Tawfik for many 
very helpful discussions. He is also very grateful to the
Pennsylvania State University Hazleton for the sabbatical leave of absence
and to the Fakult\"at f\"ur Physik  der Universit\"at Bielefeld.

\end{document}